\documentclass[twocolumn]{aastex63}

\def\dmu{${\rm pc \ cm ^ {-3}}$\,}
\def\lmu{${\rm mJy \ kpc ^ {2}}$\,}
\def\rmu{${\rm rad \ m ^ {-2}}$\,}

\def\uJybm{${\rm $\mu$Jy/beam}$\,}

\def\ga{\mbox{\raisebox{-0.1ex}{$\scriptscriptstyle \stackrel{>}{\sim}$\,}}}

\newcommand{\psrone}{J0036$-$1033}

\received{\today}
\revised{\today}
\accepted{\today}
\accepted{}
\submitjournal{ApJ}

\shorttitle{Discovery of \psrone }
\shortauthors{Swainston et al.}

\begin{document}

%
%
\title{Discovery of a steep-spectrum low-luminosity pulsar with the Murchison Widefield Array}

\correspondingauthor{N. D. R. Bhat}
\email{ramesh.bhat@curtin.edu.au}

\author[0000-0001-8982-1187]{N. A. Swainston}
\affiliation{International Centre for Radio Astronomy Research, Curtin University, Bentley, WA 6102, Australia}
\author[0000-0002-8383-5059]{N. D. R. Bhat}
\affiliation{International Centre for Radio Astronomy Research, Curtin University, Bentley, WA 6102, Australia}
\author[0000-0001-5772-338X]{M. Sokolowski}
\affiliation{International Centre for Radio Astronomy Research, Curtin University, Bentley, WA 6102, Australia}
\author[0000-0001-6114-7469]{S. J. McSweeney}
\affiliation{International Centre for Radio Astronomy Research, Curtin University, Bentley, WA 6102, Australia}
\author[0000-0002-6631-1077]{S. Kudale}
\affiliation{National Centre for Radio Astrophysics, Tata Institute of Fundamental Research, Pune 411 007, India}
\author[0000-0002-9618-2499]{S. Dai}
\affiliation{CSIRO Astronomy and Space Science, PO Box 76, Epping, NSW 1710, Australia}
\affiliation{Western Sydney University, Locked Bag 1797, Penrith South DC, NSW 1797, Australia}
\author[0000-0001-7801-9105]{K. R. Smith}
\affiliation{International Centre for Radio Astronomy Research, Curtin University, Bentley, WA 6102, Australia}
\author[0000-0003-0833-0541]{I. S. Morrison}
\affiliation{International Centre for Radio Astronomy Research, Curtin University, Bentley, WA 6102, Australia}
\author[0000-0002-7285-6348]{R. M. Shannon}
\affiliation{Centre for Astrophysics and Supercomputing, Swinburne University of Technology, P.O. Box 218, Hawthorn, VIC 3122, Australia}
\affiliation{ARC Centre of Excellence for Gravitational Wave Discovery (OzGrav), Swinburne University of Technology, Hawthorn, Australia}
\author[0000-0003-2519-7375]{W. van Straten}
\affiliation{Institute for Radio Astronomy \& Space Research, Auckland University of Technology, Private Bag 92006, Auckland 1142, New Zealand} 
\author[0000-0001-8018-1830]{M. Xue}
\affiliation{National Astronomical Observatories, Chinese Academy of Sciences, Datun Road, Chaoyang District, Beijing 100101, China}
\author[0000-0002-6380-1425]{S. M. Ord}
\affiliation{CSIRO Astronomy and Space Science, PO Box 76, Epping, NSW 1710, Australia}
\author[0000-0001-7662-2576]{S. E. Tremblay}
\affiliation{International Centre for Radio Astronomy Research, Curtin University, Bentley, WA 6102, Australia}
\author[0000-0001-8845-1225]{B. W. Meyers}
\affiliation{Department of Physics \& Astronomy, University of British Columbia, 6224 Agricultural Road, Vancouver, BC V6T 1Z1, Canada}
\author[0000-0001-9080-0105]{A. Williams}
\affiliation{International Centre for Radio Astronomy Research, Curtin University, Bentley, WA 6102, Australia}
\author[0000-0003-0134-3884]{G. Sleap}
\affiliation{International Centre for Radio Astronomy Research, Curtin University, Bentley, WA 6102, Australia} 
\author[0000-0003-2756-8301]{M.~Johnston-Hollitt}
\affiliation{International Centre for Radio Astronomy Research, Curtin University, Bentley, WA 6102, Australia}
\author[0000-0001-6295-2881]{D.~L.~Kaplan}
\affiliation{Department of Physics, University of Wisconsin--Milwaukee, Milwaukee, WI 53201, USA} 
\author[0000-0002-8195-7562]{S.~J.~Tingay}
\affiliation{International Centre for Radio Astronomy Research, Curtin University, Bentley, WA 6102, Australia} 
\author[0000-0002-6995-4131]{R.~B.~Wayth}
\affiliation{International Centre for Radio Astronomy Research, Curtin University, Bentley, WA 6102, Australia} 



\begin{abstract}

We report the discovery of the first new pulsar with the Murchison Widefield Array (MWA), PSR \psrone, a long-period (0.9 s) nonrecycled pulsar with a dispersion measure (DM) of 23.1 \dmu.
It was found after processing only
a small fraction ($\sim$1\%) of data from an ongoing all-sky pulsar survey. Follow-up observations have been made with the MWA, the upgraded Giant Metrewave Radio Telescope (uGMRT),  and the Parkes 64 m telescopes, spanning a frequency range from $\sim$150 MHz to 4 GHz.
The pulsar is faint, with an estimated flux density ($S$) of $\sim$1 mJy at 400 MHz and a spectrum $S(\nu)\,\propto\,\nu^{-2.0\pm0.2}$, where $\nu$ is frequency. 
The DM-derived distance implies that it is also a low-luminosity source ($\sim$ 0.1 \lmu  at 1400 MHz). 
The analysis of archival MWA observations reveals that the pulsar’s mean flux density varies by up to a factor of $\sim$5-6 on timescales of several weeks to months. 
By combining MWA and uGMRT data, the pulsar position was determined to arcsecond precision. 
We also report on polarization properties detected in the MWA and Parkes bands.
The pulsar's nondetection in previous pulsar and continuum imaging surveys, the observed high variability, and its detection in a small fraction of the survey data searched to date, all hint at a larger population of pulsars that await 
discovery in the southern hemisphere, with the MWA and the future low-frequency Square Kilometre Array.

\end{abstract}




\section{Introduction} \label{sec:intro}

Scanning large swathes of the sky to discover new pulsars has been an integral part of pulsar astronomy over its history. Over the past decades, multiple surveys that exploited new telescopes and/or instrumentation have led to substantial progress in understanding the pulsar population, and have uncovered a variety of exotic objects including millisecond pulsars (MSPs), sporadic emitters like rotating radio transients (RRATs), and magnetars \citep[e.g.][]{manchester2001,cordes2006,keith2010}. These discoveries have significantly advanced our understanding of the Galactic population of pulsars, besides enabling wide-ranging astrophysical studies using these objects \citep[e.g.][]{lorimer2006,levin2013}.

The advent of next-generation radio telescopes presents new avenues to explore the universe. With their wide fields of view and software-defined instrumentation, they allow us to consider novel approaches for pulsar surveys, potentially probing hitherto unexplored parts of the pulsar population parameter space. Some of these new telescopes are also important precursors or pathfinders for the upcoming Square Kilometer Array (SKA) telescope, including its low frequency (50$-$350\,MHz) component (SKA-Low). 

Over the past decade, the resurgence of low-frequency pulsar astronomy has led to new surveys undertaken by the Low Frequency Array (LOFAR), the Giant Metrewave Radio Telescope (GMRT), the Green Bank Telescope, and the Canadian Hydrogen Intensity Mapping Experiment \citep{sanidas2019,bhaswati2016,stovall2014,good2020}. These searches at frequencies below 400 MHz have already uncovered over 200 new pulsars including the fastest and slowest field pulsars known \citep{bassa2017,tan2018}, and have 
yielded 
a number of potentially promising pulsars for high-precision timing projects such as pulsar timing arrays. 
These low-frequency searches appear to be most effective in uncovering more local objects, but they require substantial computational resources to circumvent the large dispersive delays inherent at lower radio frequencies. 

Despite its modest sensitivity compared to most northern facilities such as LOFAR, the combination of a large field of view (FoV) and a superb radio-quiet environment makes the Murchison Widefield Array \citep[MWA;][]{tingay2013,wayth2018} a promising facility to undertake pulsar searches in the  southern hemisphere at frequencies below 300 MHz. The original array of 128 tiles with maximum $\sim$3 km baseline (Phase 1) was upgraded to provide  longer baselines and a flexible reconfiguration arrangement between compact (300 m) and extended arrays (Phase 2). 

Although not originally conceived or designed for pulsar science, the development of a voltage capture system (VCS; \citealp{Tremblay2015}) and associated software instrumentation \citep{Ord2019,McSweeney2020} has geared this telescope for useful pulsar work. 
The VCS allows recording of raw voltage data from a maximum of 128 tiles. 
Over the past few years the MWA has been exploited for wide-ranging pulsar science, from studies of millisecond pulsars to sporadic emission from pulsars, and from investigating the pulsar emission physics to studying propagation effects caused by the interstellar medium \citep[ISM; e.g.][]{bhat2016,mcsweeney2017,meyers2018,kaur2019}. This progress 
on both 
scientific and technical fronts to date has also paved the way to initiate a large southern-sky survey for pulsars, which will complement ongoing efforts around the world, in both sky and frequency coverage. 

Multipath propagation effects, in particular temporal broadening that becomes highly prominent at low radio frequencies, limit the detectability of short-period and distant pulsars to a few kiloparsecs in the Galactic plane \citep[e.g.][]{bhat2004}. Despite this, pulsar surveys at low frequencies tend to benefit from inherently larger fields of view (hence increased survey speeds), and to a certain extent, from the increased flux density owing to the generally steeper flux density spectrum of pulsars \citep[e.g.][]{maron2000,fabian2018}. Moreover, the much larger variability in flux densities expected at low frequencies (on timescales of weeks to months), due to diffractive and refractive scintillation effects, also increases the chances of benefiting from episodes of potential scintillation brightening, boosting the prospect of detecting relatively faint pulsars that are otherwise below the detection threshold.
However, the large sky temperatures tend to dominate the system temperatures, thereby reducing the achievable search sensitivity at low frequencies. 
A better understanding of these spectral and propagation effects may help to ascertain the detectability of pulsars with future large low-frequency facilities such as SKA-Low \citep{keane2015,xue2017}. 

\begin{figure*}
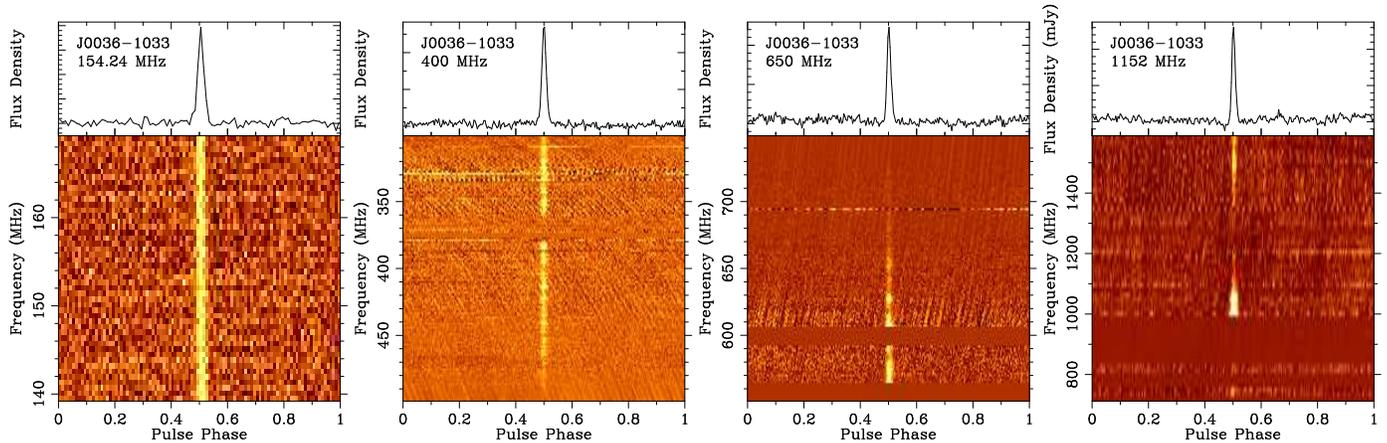

\gridline{\fig{f1a.eps}{0.25\textwidth}{}
         \fig{f1b.eps}{0.25\textwidth}{}
          \fig{f1c.eps}{0.25\textwidth}{}
          \fig{f1d.eps}{0.25\textwidth}{}
          }
\caption{
Detection plots of PSR~\psrone\ with the MWA (left panel), uGMRT (two central panels), and Parkes (right panel) telescopes, spanning a frequency range from 140 to 1600 MHz: the top panel is the integrated pulse profile  and the waterfall plot below shows the pulse strength vs. pulse phase and frequency.  MWA observations were made with the Phase 2 compact configuration of the array, whereas  those with the uGMRT made use of the 200 MHz mode of phased-array beamformer comprised of 11-13 antennas located within the central square. Parkes observations were made using the ultra-wideband low-frequency receiver (700-4032 MHz); however, the pulsar is too weak for a clear detection at $\ga$1.5 GHz.  
}
\label{fig:psrplot}
\end{figure*}

In this paper, we report on the discovery of a new pulsar with the MWA, PSR~\psrone, a nonrecycled pulsar with a period of 0.9 s and a DM of 23.1\,\dmu. It was found in the initial processing of a very small fraction ($\sim$1\%) of data from an ongoing large survey of the southern hemisphere in the 140-170 MHz band. In \S~\ref{sec:smart} and \S~\ref{sec:discovery}, we summarize observational details and data processing related to the discovery and confirmation. Our analysis of the multifrequency data is presented in \S~\ref{sec:analysis}, including evidence that the pulsar is a rare low-luminosity object. In \S~\ref{sec:discussion}, we comment on future prospects for pulsar searches with the MWA and those planned with SKA-Low, and a brief summary of the key results is given in \S~\ref{sec:conc}.

\section{The SMART pulsar survey} \label{sec:smart}

The combination of the MWA's voltage-capture mode and its recent upgrade, with a compact configuration of 128 tiles within $\sim$300 metres, allows high-survey-speed pulsar searches of the southern hemisphere at low frequencies. To take advantage of this, an all-sky pulsar search project has been initiated with the eventual goal of surveying the entire 
sky south of $+30^{\circ}$ in declination for pulsars in the 140-170 MHz band. 
This Southern-sky MWA Rapid Two-meter (SMART) pulsar survey is currently in its early stages,
with a projected timeline for data collection by $\sim$2023. 
It exploits the newly developed multipixel functionality of the MWA’s software beamformer, whereby dozens of tied-array beams are formed simultaneously (Swainston et al. in prep). 
This processing is performed on VCS data, recorded over $30.72\,$MHz at 
100-$\mu$s/10-kHz resolutions, following a two-stage channelization in the signal path, and the beam-formed output
is written out as full Stokes \citep[cf.][]{tingay2013,Tremblay2015,Ord2019}. 
Further details of the survey design, science goals, observing strategies, and processing pipelines will be described in 
a future publication (Bhat et al. in prep.).

Considering the substantial processing requirements of searching at the low frequencies of the MWA — due to large dispersion delays and the need to form thousands of beams from recorded voltage data — survey data processing is being undertaken in multiple stages: in the first pass, 10 minutes of data from each VCS observation (80 minutes) 
are 
processed and searched for pulsars, to attain a sensitivity approximately $\sqrt{10/80} \sim 1/3$ of that which would be eventually attainable, a strategy designed to boost the prospects of early discoveries. Each 10 minute observation is processed to generate $\sim$6300 beams 
that tessellate 
the $\sim500\,{\rm deg}^2$ field of view at $\sim$ 150 MHz. Each tied-array beam (at 100-$\mu$s/10-kHz resolutions) is searched out to a DM of 250 \dmu, in 2358 DM trials, with down sampling by up to a factor of 16 (i.e.\ time resolution of 0.1 to 1.6 ms). 
At this stage, our pulsar-detection algorithms are limited to periodicity and single-pulse searches,  
(i.e. no acceleration searches). 
The search part of the processing chain makes use of the standard suite of tools in the {\tt PRESTO} software package \citep{presto} and the machine learning classifier that was developed for the LOFAR pulsar survey \citep{tan2018a}. The calibration performance, data quality, and 
analysis pipeline are verified 
via the detection of known pulsars in the field. The software is run on the Swinburne’s OzSTAR supercomputer.

\begin{figure*}[t]
\vspace{-0.20in}
\gridline{\fig{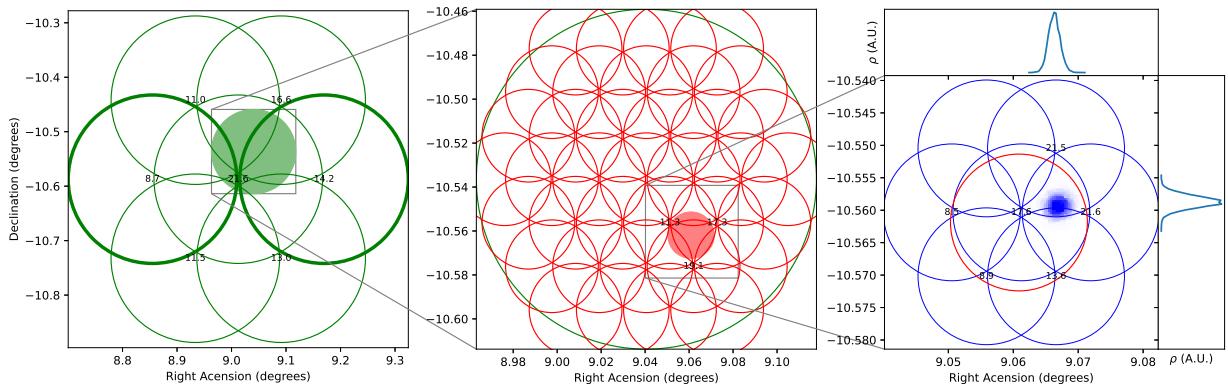}{1.1\textwidth}{}}
\vskip -1.0cm
\caption{
Positional determination of the pulsar via tied-array beam localization by reprocessing VCS observations. The pulsar was first detected in two compact array beams ($\sim$20$^{\prime}$; thick green circles), the sky region around which was subsequently covered with a hexagonal grid of beams (thinner green circles). A denser grid was subsequently made with the Phase 1 array beams ($\sim$2$^{\prime}$; red circles), the best position from which (the red circle on the right panel) was then subjected to a dense hexagonal grid using data from the Phase 2 extended configuration ($\sim$1$^{\prime}$; blue circles). These latter detections were subjected to the method in Bannister et al. (2017) to derive a probabilistic distribution (the bluish region in the right panel), where the positional uncertainties are $\sim$12$^{''}$, approximately one-tenth of the tied-array beam size at 140-170 MHz, with the extended configuration. 
}
\label{fig:beam_localization}
\end{figure*}

\begin{deluxetable*}{lccccc}[t]
\tablecaption{Summary of Follow-up Observations \label{tab:obs}}
\tablecolumns{7}
\tablenum{1}
\tablewidth{0pt}
\tablehead{
\colhead{Telescope/Receiver} &
\colhead{Frequency Range (MHz)} &
\colhead{MJD Range } &
\colhead{No. of Spectral Channels} &
\colhead{Time Resolution ($\mu$s)} &
\colhead{No. of Epochs} 
}
\startdata
MWA VCS & 140 - 170  & 57366 - 59178 & 3072 & 100  & 16 \\
uGMRT Band 3 & 300 - 500  & 59158 - 59184 &  2048 & 655.36 & 3 \\
uGMRT Band 4  & 550 - 750  & 59159 - 59188 & 2048 & 655.36 & 2 \\
Parkes UWL & 704 - 4032  & 59132 - 59171 & 3328 & 256 & 6 \\
\enddata
\end{deluxetable*}

\section{Discovery and Followup Observations} \label{sec:discovery}

PSR \psrone\ is the first new pulsar 
discovered during 
our ongoing processing. It has a period of 0.9 s and 
a DM of 
23.1$\pm$0.2\,\dmu,
and was first detected with a signal-to-noise ratio (S/N) $\sim$ 10-12  in observations taken on MJD 58774. Subsequently, the full 80 minute observation (42 TB) was processed and 
reanalysed to yield a much improved detection with S/N $\sim$ 36, thereby providing unambiguous initial confirmation of the candidate (Fig.~\ref{fig:psrplot}). The fact that the raw voltage data were recorded (instead of the more common filterbank data typically employed in most pulsar searches), along with access to archival VCS data and the suite of available \textit{post hoc} interferometric/tied-array processing options, allowed a multitude of important follow-up analyses to (1) confirm the discovery, (2) obtain a precise sky localisation, and (3) undertake initial polarimetric and variability studies of the pulsar. 
All these analyses were performed on beamformed data at 100-$\mu$s/10-kHz resolutions, and made use of incoherent dedispersion, resulting in a residual dispersive  smearing of 0.7 ms, which is much less than  the measured pulse width of $\sim$20 ms (Fig.~\ref{fig:psrplot}). 
Subsequently, new observations commenced in 2020 June for timing and imaging follow-ups, the analysis and results of which are summarized in the following sections.

\subsection{Follow-up observations} \label{sec:follow-up}

Besides the MWA, follow-up observations were also made using the uGMRT and Parkes telescopes, the details of which are summarized in Table~\ref{tab:obs}. The pulsar was detected at 
two to three epochs, 
separated by $\sim$one to three weeks with both the uGMRT and Parkes, and a total of 16 detections were made with the MWA over a time span of five months.

\subsubsection{The MWA} \label{sec:mwa}


The MWA  operated in the extended configuration from 2020 May, and VCS data were recorded over a duration of $\sim$20 to 60 minutes. 
These data were processed to generate beamformed time series, and were analyzed using the standard pulsar packages such as {\tt DSPSR} \citep{dspsr} and {\tt PSRCHIVE} \citep{psrchive}. Unlike the case with the data from Parkes and the uGMRT, MWA data are generally devoid of radio-frequency interference (RFI; e.g., Fig.~\ref{fig:psrplot}), and hence 
do not require any specific procedures for excising RFI.  
%
%

\begin{figure*}
\includegraphics[width=1.70in,angle=270]{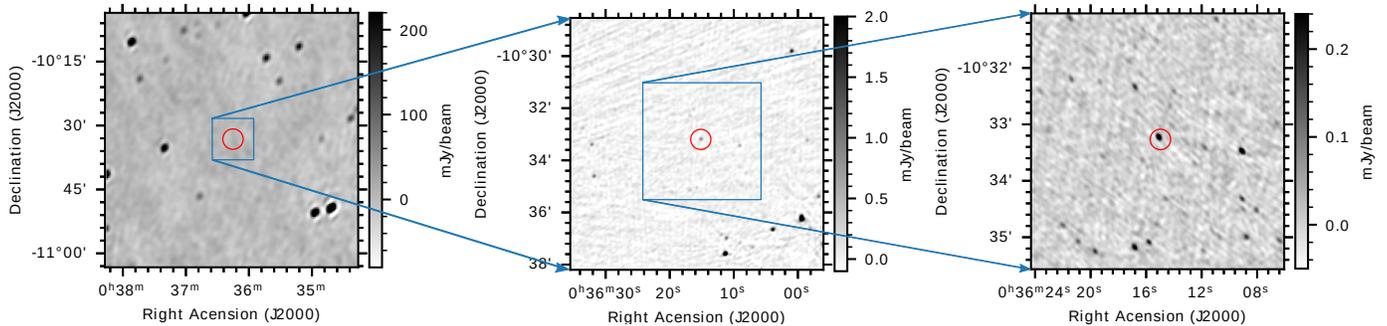}
\caption{
{\it Left:} MWA Stokes I image of the $1\times1$\,$\text{deg}^2$ field (at 154.88 MHz) with the position of the target pulsar in the center (indicated by a circle of the radius $\approx$5 times the position error of $\approx$12\,arcsec); the rms sensitivity is $\approx$8 mJy/beam.  {\it Middle:} uGMRT image in Band 3 (300-500 MHz) where the rms sensitivity is $\approx$156 $\mu$Jy/beam, yielding a clear $\sim$6-$\sigma$ detection of the pulsar. {\it Right:} uGMRT image in Band 4 (550-750 MHz), where a much higher sensitivity of $\approx$15 $\mu$Jy/beam was achieved, resulting in a $\sim$35-$\sigma$ detection of the pulsar. The detections in both images are marked by red circles.
The estimated positions have about $\sim$1 arcsecond uncertainties and they differ by approximately 4 arcseconds, which may arise from residual errors from the ionospheric calibration.}
\label{fig:imaging}
\end{figure*}

\subsubsection{The uGMRT} \label{sec:gmrt}

Observations with the upgraded GMRT \citep[uGMRT;][]{ugmrt} were made concurrently in the imaging and 
phased-array 
modes, i.e.\ using the wide-band correlator \citep{gwb} in conjunction with the phased-array backend 
that generates beam-formed (and channelized) data. These observations were made in both Band 3 (300-500 MHz) and Band 4 (550-750 MHz). 
Due to $\sim1^\prime$ positional uncertainty from the initial localization that was achieved 
with the MWA
(cf. \S~\ref{sec:localization}), two phased-array beams were formed: Beam 1 consisted of antennas located within the central square (typically $\sim$12), whereas two additional antennas from each of the three arms were added to all working antennas from the central square to form Beam 2. These data were recorded 
as 
a 2048-channel spectrum every $655.36\,\mu$s. 
Examples of phased-array beam detections (Bands 3 and 4) are shown in 
Fig.~\ref{fig:psrplot}.

The visibility data were recorded at the standard 2.68-second resolution. Each observation of the target pulsar was preceded by observations of the phase calibrator 0025$-$260, which is sufficiently 
bright ($S_{325}\,\sim$\,20\,Jy) to perform bandpass calibration. Prior to starting each cycle of target observation, the array was rephased, and hence, two observations of the phase calibrator (i.e.\ immediately preceding and following the scan) were available for calibration purposes. An automated imaging pipeline (Kudale et al. in prep.), composed of {\tt flagcal} \citep[Chengalur 2013;][]{flagcal}, {\tt PyBDSM} \citep{pybdsm}, and 
the Common Astronomy Software Applications package
\citep[CASA;][]{casa}, was used for continuum analysis.  In total, three self-calibration and imaging cycles (two phase only, and one amplitude and phase) were carried out. These self-calibrated visibility data were then imaged, in four subbands that span the 200-MHz bandwidth, by choosing an appropriate range of channels that span approximately 48 MHz per subband.

\subsubsection{Parkes} \label{sec:parkes}

Observations with the Parkes 64 m telescope (recently given the indigenous Wiradjuri name {\it Murriyang})
made use of the the Medusa backend, in conjunction with the ultra-wideband low-frequency (UWL) receiver, which provides
instantaneous 
coverage from 704 to 4032 MHz 
\citep[for details, see][]{hobbs2020}. 
Data were recorded with two-bit sampling every 256 $\mu$s in each of the 1-MHz wide frequency channels (3328 channels in total). With the spin period and DM determined by MWA observations, Parkes data were folded using the {\tt DSPSR}  software package with a subintegration length of 30 s. We manually excised data affected by narrowband and impulsive radio-frequency interference for each subintegration. The best detection from our Parkes observations to date is shown in Fig.~\ref{fig:psrplot}. 
Polarization and flux density calibration were carried out using {\tt PSRCHIVE}  and followed procedures described in \citet{dai2019}.

\section{Analysis and Results} \label{sec:analysis}

\subsection{Localization} \label{sec:localization}

The positional uncertainty of the original detection was $\sim 20^\prime$ (the size of the Phase 2 compact tied-array beam). However, precise localization of the pulsar's position to the level of a few arcseconds (i.e.\ more than two orders of magnitude improvement) was possible thanks to the subsequent detection of the pulsar in archival MWA data spanning Phase 1, Phase 2 compact, and Phase 2 extended configurations, coupled with the reprocessing flexibility afforded by the VCS. The essential methodology is depicted in Fig.~\ref{fig:beam_localization}, and briefly summarized below. Subsequent resolution of systematic errors present in the MWA localization measurements was achieved via high-resolution imaging of the pulsar field with the uGMRT.

\subsubsection{Tied-array beamforming with the MWA} \label{sec:tiedarraybeams}

Following the original detection using the Phase 2 compact array 
(on MJD 58774),
VCS observations containing the pulsar were reprocessed to make a hexagonal grid of six pointings around the beam in which the pulsar was detected. These detections (corresponding S/Ns) were then used to estimate a maximum likelihood position by assuming Gaussian-shaped beam patterns (with full width at half maximum, FWHM $\sim$19.5’) for the central part of the main lobe of the tied-array beam pattern. 
An archival VCS observation that covered the pulsar position (on MJD 57406 using the Phase 1 array) was then reprocessed to make a denser grid of three hexagonal rings around the revised position.  The resulting detections were then used to obtain a much-improved maximum likelihood position estimate, made possible by the substantially narrower tied-array beams (FWHM $\sim$2.5’).
This was then further improved to subarcminute 
precision 
using new observations taken in the extended configuration, which were processed to make a hexagonal grid of six pointings (Fig.~\ref{fig:beam_localization}), where tied-array beams are even smaller (FWHM $\sim$1.2') owing to longer baselines. 

These detections were also analyzed using the localization strategy routinely used for initial Fast Radio Burst localization with the Australian Square Kilometre Array Pathfinder \citep{bannister2017,shannon2018}, to obtain a more robust position, where statistical uncertainties are $\sim$12” (in both RA and DEC). For this final step, we employed a more accurate beam size, based on the point spread function 
(PSF)
obtained from imaging of the corresponding VCS data. The best position from MWA data (J2000) was RA = 00:36:16, DEC = $-$10:33:32. However, applying this method to multiple observations revealed a positional wandering of $\sim30$-$40^{\prime\prime}$, 
possibly due to a combination of ionospheric and calibration effects. Accurate localization was ultimately achieved via uGMRT high-resolution imaging. 

\subsubsection{High-resolution imaging with the uGMRT} \label{sec:gmrtimaging}

Following the initial localization via the MWA tied-array beam processing, sensitive high-resolution imaging analysis was undertaken with uGMRT Band 3 and Band 4 observations. Images of the fields around the pulsar position are shown in Fig.~\ref{fig:imaging}, where $\sim 8^\prime \times 8^\prime$ and $\sim 5^\prime \times 5^\prime$ regions around the pulsar position are shown. With $\sim$25 km maximum baselines, the size of the PSF is approximately $10^{\prime\prime}$ and $5^{\prime\prime}$, respectively in Bands 3 and 4, thus enabling high-resolution imaging of the field. The high sensitivity of the uGMRT 
enabled subband imaging 
(wherein the 200 MHz observing band was split into $4 \times 50\,$MHz chunks). The rms varied from $\sim$71 to 223 $\uJybm$ across Band 3, while Band 4 imaging reached a much lower rms of $\sim$40 to 65 $\uJybm$. The best image (from Band 4 imaging on 2020 December 4) reached an rms of $\sim$15 $\uJybm$ at 650 MHz, resulting in a high-fidelity image of the field and a $\sim$35-$\sigma$ detection of the pulsar. The resultant positions are: RA = 00:36:15.15 and DEC = $-$10:33:11.9 (Band 3), and RA = 00:36:14.97 and DEC = $-$10:33:16.02 (Band 4), with uncertainties $\sim0.5^{\prime\prime}$ and $\sim0.1^{\prime\prime}$ in Bands 3 and 4, respectively. The $\approx 4^{\prime\prime}$ discrepancy in the positions can be attributed to residual errors from ionospheric calibration, which impacts Band 3 calibration more, and we therefore quote a mean position with $\sim5^{\prime\prime}$ uncertainties,
i.e.\ RA = 00:36:15.1, DEC = $-$10:33:14, which correspond to
Galactic coordinates $l = 112.3^{\circ}$ and $b = -72.9^{\circ}$. 

\begin{deluxetable}{ll}[b]
\tablecaption{Parameter Summary of PSR~\psrone \label{tab:timing}}
\tablecolumns{2}
\tablenum{2}
\tablewidth{0pt}
\tablehead{
\colhead{Parameter} &
\colhead{Value}
} 
\startdata
Right ascension (J2000) & 00$^h$36$^m$15$^s$.01(4) \\
Declination (J2000)     &  $-$10$^o$33$^{\prime}$14$^{''}$.2(9)  \\
Galactic longitude ($l$) & $112.3^{\circ}$ \\
Galactic latitude ($b$) & $-72.9^{\circ}$ \\
Epoch (MJD) & 58774 \\
Spin period ($P$) & 0.900009289(3)\,s \\
Period derivative ($\dot{P}$) & $2.131(3) \times 10^{-16}\,{\rm s\,s^{-1} }$ \\
Dispersion measure (DM) & $23.1\pm0.2$\,\dmu \\
Flux density at 400 MHz & 1\,mJy\\
Rotation measure (RM) & $-8.1\pm0.7$\,\rmu \\
Surface magnetic field ($B$) & $4.4 \times 10^{11}$\,G \\
Characteristic age ($\tau _c$) & 67\,Myr \\
\enddata
\label{tab:timing}
\end{deluxetable}

\subsection{Imaging with the MWA} \label{sec:imaging}

The recorded VCS data can also be correlated offline and imaged. Coadding the initial two $\sim$20-30 minute observations achieved only a sensitivity of $\sim$6-8 mJy/beam for Stokes I, yielding a nondetection of any source near the expected position. Coadding a further four more observations yielded only a marginal improvement in sensitivity. There appears to be a tantalizing source near the uGMRT localized position (Fig.~\ref{fig:imaging}), with a flux density $\sim$16\,mJy; however the significance is too low 
($\sim$2$\sigma$) 
to claim a confident positive detection. 

\begin{figure*}
\begin{center}
\includegraphics[height=160mm,angle=270]{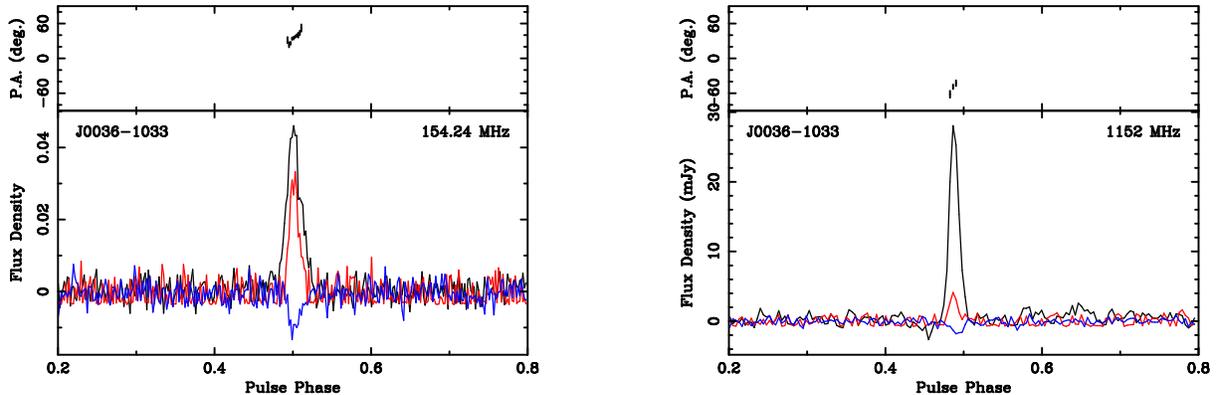}
\end{center}
\caption{Polarimetric profiles of PSR J0036-1033 at MWA 155 MHz (left) and at Parkes 1.1 GHz (right).
The black, red, and blue curves in the lower panels show the total intensity, linear, and circular polarization, respectively.
An analysis of the PA curve (MWA) yields 
$d \psi / d \phi = {\rm sin} \alpha / {\rm sin} \beta = 3.8$.
The pulsar is weakly polarized at 1.1 GHz, 
and RM = $-8.1 \pm 0.7$ \rmu from MWA data. 
The flux density scale is in arbitrary units for MWA data, 
for which absolute flux density calibration is not possible. 
}
\label{pol}
\end{figure*}

\subsection{Timing} \label{sec:timing}

Timing analysis of recent data (MJD = 59000 - 59150) yielded an initial solution for the basic parameters, i.e. the spin frequency
and its first derivative, which were 
subsequently refined with the addition of times of arrival from past observations (Fig.~\ref{fig:variability}).
We used the GMRT-determined position as an initial estimate, and then progressively improved the timing solution, with the gradual addition of data from the past years. The timing solution from this 5-yr time span is in Table~\ref{tab:timing}. An inferred magnetic field of $B$
$\sim$ $4.4 \times 10^{11}$\,G and a characteristic age, $\tau _c$
$\sim$ 67\,Myr confirms that it is a nonrecycled pulsar. 

\subsection{Polarimetry}

Polarimetric data were available for both the MWA and Parkes observations.
The beamformed MWA data were obtained by the system described in \citet{Ord2019} and \citet{Xue2019}.
The Faraday rotation measure synthesis technique \citep{Brentjens2005} was applied to one of the brightest MWA (MJD 58774.6) detections, yielding an estimated rotation measure (RM) of $-8.1 \pm 0.7$\,\rmu.

After correcting for Faraday rotation, linear and circular polarization was detected, with the fractional polarization being higher at the lower frequency (Fig. \ref{pol}). The Parkes system, which employs noise diodes for polarimetric calibration, has been well tested and verified to yield accurate measurements \citep[e.g.][]{hobbs2020}. The qualitative agreement between the profiles at the two frequencies (the slope of the position angle, PA, 
and the sign of the circular) indicates that the MWA polarimetric profiles are at least approximately correct \citep[see][]{Xue2019a}, while the decrease in the fractional linear polarization at higher frequencies is a common feature of pulsar emission \citep{Manchester1973}.

We attempted to fit the rotating vector model \citep{Radhakrishnan1969} to the 
PA of the linear polarization 
across the on-pulse window, in order to constrain the viewing geometry, $(\alpha, \beta)$, where $\alpha$ is the angle between the magnetic and rotation axes and $\beta$ is the impact angle of the magnetic axis on the line of sight. In the absence of relativistic effects, the PA curve is expected to be steepest in the center of the pulse profile, with slope $d\psi/d\phi = \sin\alpha / \sin\beta$, where $\psi$ is the PA at phase $\phi$. Across all observations for which there was sufficiently high S/N, the measured slope was consistent with a value of $d\psi/d\phi \sim 3.8\,^\circ/^\circ$. Further constraints on the viewing geometry by this method are unlikely unless the PA can be measured over a wider range of phases (for example, if a broader profile is measured at even lower frequencies).

\subsection{Archival detections}

The fortuitous availability of multiple archival VCS observations that covered the pulsar position and the MWA's operational strategy to archive VCS data for future processing have also enabled multiple ($\sim$10) redetections of the pulsar in archival data. Many of these are 
a result of coincidental proximity to 
the well-known subpulse drifting pulsar J0034$-$0721, which has been the subject of recent 
study using the MWA
\citep{mcsweeney2017,mcsweeney2019}. A few  observations were made with the original Phase 1 array, most with the compact Phase 2 array, and just one with the extended (Phase 2) array.  
%
The amount of time for which \psrone\ was in the primary beam for these observations varied
from $\sim$10 to $\sim$80 minutes, and the pulsar position offset from the primary beam center from $\sim$3$^{\circ}$ to $\sim$12$^{\circ}$. The data were processed to generate tied-array beams at the  best determined position as described in \S~\ref{sec:localization}, resulting in positive detections of the pulsar in nearly all observations. Almost all these detections have S/N $\sim$10-20,  which makes the discovery observation (on MJD 58774), where S/N exceeded $\sim$30, an exceptional one. Nearly half of these detections were at $\sim$8-10$^{\circ}$ offset from the primary beam center, i.e.\ near half power points of the primary beam, thus incurring a loss of S/N by a factor of two. 

\begin{figure*}
\gridline{\fig{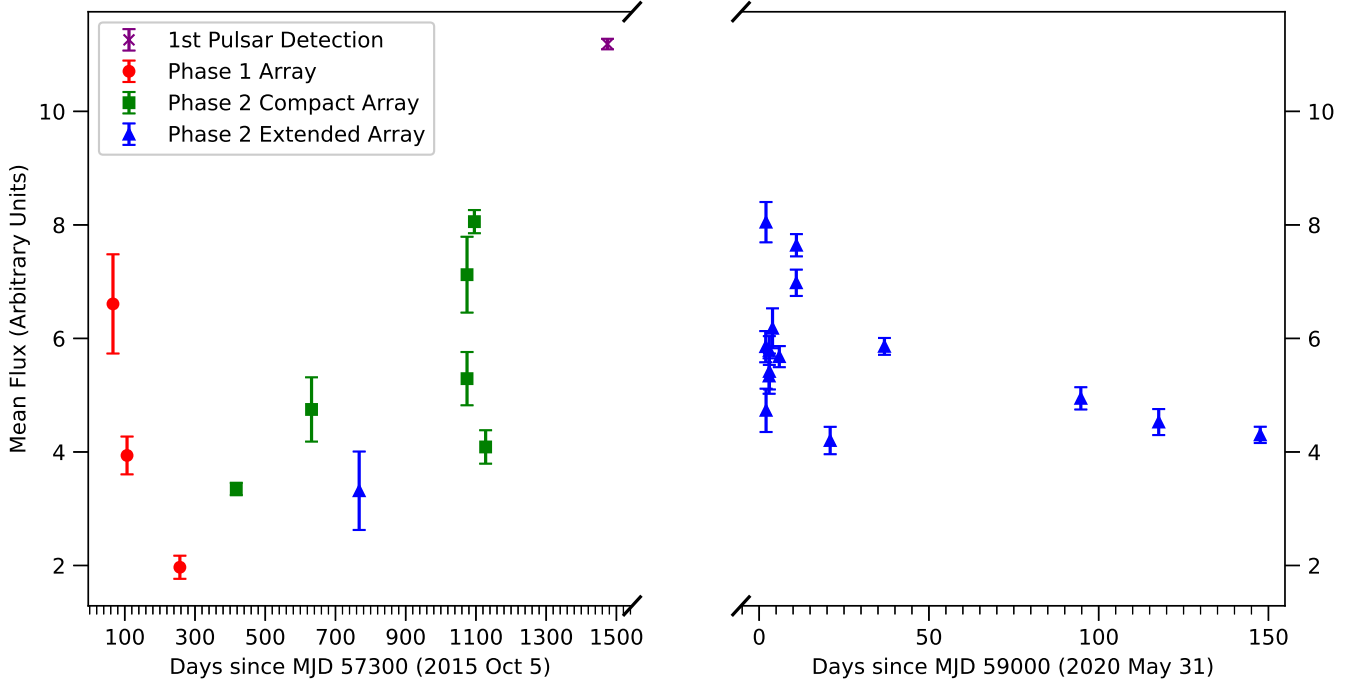}{1.0\textwidth}{}}
\vskip -1.0cm
\caption{
Variability in mean flux density of PSR \psrone, over a 4 yr time span; the quantity plotted is indicative of the mean flux density of the pulsar, after applying corrections for the primary beam, and normalized to the zenith gain, and fixed integration (20 minutes) and 30 MHz bandwidth. The data shown on the left are from archival observations dating back to 2016, whereas those on the right segment are from new observations made (since 2020 June) with the extended configuration of the array.  
}
\label{fig:variability}
\end{figure*}

A summary of these detections is shown in Fig.~\ref{fig:variability}, along with those from new observations made with the extended configuration in which the MWA has been operating since 2020 May. The quantity plotted here is essentially a mean flux density equivalent, i.e.\ the total pulsar flux density after normalizing for the zenith gain and scaling for constant integration time (20 minutes) and 30.72 MHz bandwidth. 
As evident from the figure, the pulsar displays significant variability, by a factor of as much as $\sim$5-6 over a $\sim$4 yr time span, further discussion of which is deferred to \S~\ref{sec:discussion}. 

\begin{figure}
\begin{center}
\includegraphics[height=80mm,angle=0]{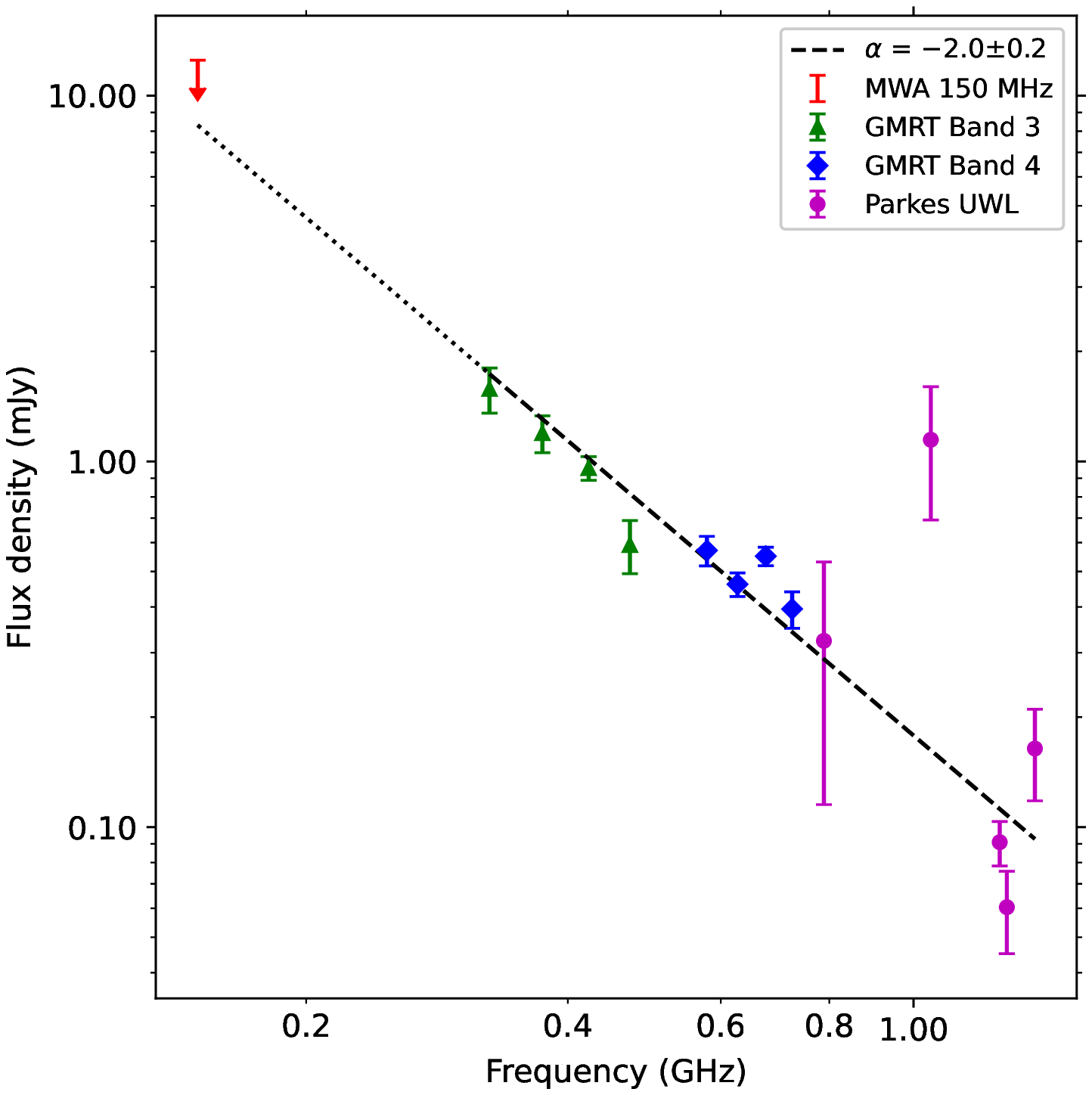}
\end{center}
\vspace{-0.5cm}
\caption{Flux density measurements of PSR~\psrone\ over a decadal frequency range from $\sim$150 to $\sim$1500 MHz. For uGMRT measurements (300-750 MHz) the uncertainties are dominated by variability between the observing epochs, whereas for Parkes the dominant source of uncertainty is the large spectral modulation seen across the 900 MHz band. A weighted least squares fit yields $\alpha = -2.0 \pm 0.2$. Excluding the apparent outlier measurement at 1.1 GHz (where the pulsar was unusually bright; cf. Fig.~\ref{fig:psrplot}) makes negligible difference to the fit. No detection was made in continuum imaging with the MWA;
however, the resultant limit is consistent with the estimated spectral slope. 
}
\label{fig:fluxdensities}
\end{figure}

\subsection{Flux densities and spectral index}

The spectral index $\alpha$, defined as $S \propto \nu^{\alpha}$, where $S$ denotes the flux density at frequency $\nu$, is estimated from uGMRT and Parkes measurements that span a frequency range from $\sim$300 to 1500 MHz. 
For uGMRT data, $S (\nu)$ was estimated by fitting a 2D Gaussian to the pulsar image using the {\tt imfit} task in CASA, and the uncertainties were estimated as the rms near the pulsar location using the {\tt imstat} task in CASA.
For Parkes data, observations of Hydra A, along with those of the pulsed noise signal, were used to derive flux density scales, using the procedure described in \citet{dai2019}.  

Fig.~\ref{fig:fluxdensities} shows a summary of the related measurements and analysis. For the uGMRT bands, which span 300-750 MHz, two independent observations separated by $\sim$3 weeks were used, thus providing reasonable values for mean flux densities that account for variability arising from scintillation effects. Parkes detections are, however, from observations separated by a few days, where  the main source of uncertainty is the substantial spectral modulation seen in one of our  observations (Fig.~\ref{fig:psrplot}). Furthermore, with the large fractional bandwidths (50\% in uGMRT Band 3, 30\% in Band 4, and 70\% in Parkes UWL), we are able to make multiple measurements within each of the bands, which were split into a suitable number of subbands based on the expected values for the scintillation bandwidth ($\nu_d$). 

The estimates of $\nu_d$ are $\sim$10 kHz in the MWA band (150 MHz), and $\sim$0.5 MHz and $\sim$7 MHz, respectively, in the uGMRT bands centered at 400 and 650 MHz, whereas a much larger value of $\sim$110 MHz may be expected at Parkes 1.2 GHz. Since the pulsar is too weak for scintillation analysis, these are first order estimates based on our MWA observations of PSR J0437$-$4715 \citep{bhat2018}, a relatively high Galactic latitude pulsar with a low DM (2.65 \dmu), and assuming the theoretical scaling in DM for the scintillation bandwidth ($\nu_d \propto DM^{-2.2}$; \citealp{cwb85}). We therefore split the uGMRT 200 MHz range into 4 $\times$ 50 MHz subbands, whereas the Parkes band was split up into three uneven bands of 200, 300 and 400 MHz to cover the 900 MHz range (700-1600 MHz) in which the pulsar detection was made. Clearly, relatively smaller values of $\nu_d$ (and hence several to many scintles) mean short-term (diffractive) scintillation effects are sufficiently averaged over in both the uGMRT bands, albeit to a lesser extent in the Parkes band. Within the constraints and quality of these measurements, we estimate a spectral index of $\alpha = -2.0 \pm 0.2$ (see Fig.~\ref{fig:fluxdensities}).

\section{Discussion} \label{sec:discussion}

A striking aspect of the new pulsar is the indication of a spectrum steeper than most long-period pulsars (Fig.~\ref{fig:fluxdensities}), consistent with the average spectral index of 21 pulsars discovered with LOFAR \citep{Tan2020}. However, in light of a the substantial variability observed for this pulsar (Fig.~\ref{fig:variability}), it is possible that our current estimates of mean flux densities, and hence the spectral index, may depart from their true values. Our inferred value, $\alpha \approx -2$, is largely based on the measurements made in the uGMRT bands (300-700 MHz) and part of the Parkes band (700-1500 MHz), where the estimated flux densities are more reliable. The nondetection of the pulsar in the continuum imaging in the MWA band (with the implied constraint of $S_{150} \lesssim 10\,$mJy) is consistent with the extrapolation of the flux density to MWA frequencies, even assuming no spectral turnover.

The NE2001 model of \citet{ne2001} gives a distance estimate of 1.06 kpc for the pulsar. 
The YMW16 model \citep{ymw16} is, howeve, unable to constrain the distance, as DM saturates at 20.38\,\dmu\ ($\sim 6\,$kpc) toward the pulsar's line of sight ($b = -72.9^{\circ}$); whereas for NE2001, DM saturates at a larger value of 30.61\,\dmu ($\sim 9\,$kpc). 
Since the YMW16 model is generally less reliable than NE2001 for pulsars at high Galactic latitudes, we use the NE2001 model for the distance and luminosity estimates.
The spectral fit (Fig.~\ref{fig:fluxdensities}), coupled with a distance estimate of 1.06 kpc, implies a luminosity $L_{1400}$ $\sim$0.1 \lmu at 1400 MHz. 
With only $\sim$40 pulsars (out of 2900) known with $L_{1400}$ below this value, this places \psrone\ in the lowermost $\sim$2\% of the currently known population of long-period pulsars.

The steep spectrum and low luminosity, along with the pulsar's high variability,  likely prevented the detection of the pulsar in previous pulsar and continuum imaging surveys. The pulsar was well below the limiting sensitivities of both the Parkes high time resolution universe survey \citep[$\sim0.3-0.6\,$mJy at 1400 MHz;][]{keith2010} and the Parkes southern pulsar survey \citep[$\sim 3\,$mJy at 430 MHz;][]{70cm}, toward high Galactic latitudes. The pulsar was also not detected in the TIFR GMRT Sky Survey \citep{tgss2017} that scanned the entire sky at declination $\delta > -53^{\circ}$ at 150 MHz, down to a sensitivity of $\sim$2-3 mJy/beam, i.e.\ over an order-of-magnitude deeper than the MWA GLEAM survey \citep{gleam2017}. On the other hand, the original MWA discovery was likely facilitated by the pulsar's high variability. As evident in Fig.~\ref{fig:variability}, the pulsar was several times brighter than its average when the first detection was made on MJD 58774, with S/N $\sim$10 in the initial search, for which we may expect a limiting sensitivity $S_{150} \gtrsim 10\,$mJy, based on our processing  parameters. The detection is thus an excellent demonstration of the efficacy of surveying at low radio frequencies, where scintillation brightening can potentially result in the detectability of objects that are otherwise well below the sensitivity limits of searches. 

As such, this discovery heralds a potentially rich harvest of new pulsar discoveries to be made in the remaining shallow pass of the SMART survey, as well as the planned deep survey. To estimate this yield, we have attempted first-order simulations of the survey, using the formalism outlined in \citet{xue2017}, based on the simulation software {\tt PsrPopPy} \citep{Bates2014}. 
The simulations take into account the sky dependence of the system temperatures at these low frequencies as well as the loss in the array gain at large zenith angles.
With the caveat that our understanding of the pulsar luminosity function and beaming fraction is limited, we project the deep survey to reach a limiting sensitivity of $\sim$2-3 mJy, with a potential net yield of $310 \pm 100$ new pulsars. 
This projection applies to the population of long-period pulsars and does not account for other classes of pulsars such as sporadic emitters (e.g. RRATs), or millisecond and binary pulsars, whose populations are hard to model or simulate. Indeed, the detection of millisecond and binary classes also require high-sensitivity and acceleration searches, which are currently outside the scope of our initial processing.
Further details on the survey yield analysis will be presented in the survey description paper (Bhat et al. in prep.). 

Assuming an isotropic distribution of our simulated local pulsar population (DM$\lesssim 250$ \dmu), and scaling for the current search sensitivity (one-third of the full search sensitivity) and the sky coverage of observations searched to date ($\sim$5\% of the sky visible to the MWA), we expect $\sim$1-2 pulsars from this initial shallow search. The detection of one new pulsar is thus in line with this general expectation. While this may seem fortuitous, we argue that the unique advantages of an MWA pulsar survey, especially from a southern hemisphere, radio-quiet environment, along with our survey parameters (e.g. long dwell times), offer excellent prospects of new pulsar discoveries, despite the MWA’s modest sensitivity and substantial processing challenges. 

The analysis presented in this paper also demonstrates several benefits of our pulsar survey. The most distinct feature of 
which
is the use of the VCS, which records the raw voltage data from individual tiles. While such a strategy poses considerable computational challenges and overheads, it also offers several unique benefits; most notably, the flexibility to reprocess the full observation to confirm the candidate, perform an initial polarimetric analysis, and redetect the pulsar in archival observations spanning several years, as far back as 2016, when early science was undertaken using the VCS capability. Without this latter ability, the highly variable nature of the pulsar, on timescales of several weeks to months, would not have been apparent.

Furthermore, access to data from all three different array configurations of the MWA (i.e.\ Phase 1, Phase 2 compact, and Phase 2 extended) has allowed us to attain a progressively improved localization, from an initial $\sim20^\prime$ 
to $\sim10^{\prime\prime}$, i.e. 
more than two orders of magnitude improvement. This enabled detailed follow-ups with more sensitive telescopes such as uGMRT and Parkes at higher frequencies, which in turn helped establish the possible steep-spectrum, low-luminosity nature of the pulsar. It also highlights the fact that even telescopes of modest sensitivity (such as the MWA) can be meaningfully leveraged with more sensitive facilities with common sky visibility, for pulsar searching applications. Since the MWA is also a precursor for SKA-Low, this also demonstrates the excellent prospects of using SKA1-Low in conjunction with SKA1-Mid or other sensitive high-frequency telescopes in order to undertake detailed follow-up studies of new pulsar discoveries that are forecast at the low frequencies of SKA-Low.

\section{Summary and Conclusions} \label{sec:conc}

With the discovery of a new pulsar, PSR~\psrone, from searching a mere $\sim$1\% of the survey data collected to date,
we have demonstrated the efficacy and relevance of low-frequency pulsar searches in the southern hemisphere. This marks an important milestone for pulsar searches with the MWA, notwithstanding its modest sensitivity and the processing and data management challenges arising from nontraditional pulsar instrumentation. The pulsar is seemingly unremarkable in terms of profile and emission characteristics, but appears to have a relatively steep spectrum ($\alpha \approx -2$) and low luminosity ($L_{1400}\,\sim\,0.1\,$\lmu), and exhibits substantial variability in flux density. It reaffirms the importance of low-frequency pulsar surveys for probing the low-luminosity population of pulsars, whereas a nondetection in previous continuum imaging and pulsar surveys underscores the benefits of exploring new parameter space as well as opportunistic discoveries facilitated by favorable episodes of scintillation brightening. 

The pulsar was promptly followed up with 
the uGMRT and Parkes, besides the MWA. This demonstrates how instruments with common-sky visibilities can be fruitfully leveraged for detailed characterization of new pulsars, including improved positional determination and broadband high-frequency follow-ups. For the first time, the archival voltage (VCS) data from the MWA was extensively exploited for confirmation and localization of a new pulsar, and to investigate its time variability. A notable highlight is the progression from initial $\sim20^\prime$ uncertainty in sky localization to $\sim10^{\prime\prime}$ through suitable reprocessing of archival data, and eventually to a few arcsecond precision via high-resolution imaging with the uGMRT. 
This discovery hints at a promising future for pulsar searches with the MWA, as the
capability to perform full-sensitivity searches (e.g. using coherent dedispersion and acceleration searches) becomes available. 
It also foreshadows a potentially rich yield for
future low-frequency pulsar surveys planned with SKA-Low. 

\acknowledgments

We would like to thank the referee for several useful comments that improved the presentation and clarity of this paper. 
This scientific work makes use of the Murchison Radio-astronomy Observatory, operated by CSIRO. We acknowledge the Wajarri Yamatji people as the traditional owners of the Observatory site. This work was supported by resources provided by the Pawsey Supercomputing Centre with funding from the Australian Government and the Government of Western Australia.
This work was supported by resources awarded under Astronomy Australia Ltd's ASTAC merit allocation scheme on the OzSTAR national facility at the Swinburne University of Technology. The OzSTAR program receives funding in part from the Astronomy National Collaborative Research Infrastructure Strategy (NCRIS) allocation provided by the Australian Government.
The GMRT is run by the National Centre for Radio Astrophysics
of the Tata Institute of Fundamental Research, India. 
The Parkes radio telescope is part of the Australia Telescope National Facility which is funded by the Australian Government for operation as a National Facility managed by CSIRO.
RMS receives support through Australian Research Council Future Fellowship FT190100155 and CE170100004.
DK acknowledges the  NANOGrav Physics Frontiers Center through NSF awards 1430284 and AST-1816492.
\software{CASA \citep{casa}, DSPSR \citep{dspsr}, PRESTO \citep{presto}, PSRCHIVE \citep{psrchive}, PsrPopPy \citep{Bates2014}}


\begin{thebibliography}{}
\expandafter\ifx\csname natexlab\endcsname\relax\def\natexlab#1{#1}\fi
\providecommand{\url}[1]{\href{#1}{#1}}
\providecommand{\dodoi}[1]{doi:~\href{http://doi.org/#1}{\nolinkurl{#1}}}
\providecommand{\doeprint}[1]{\href{http://ascl.net/#1}{\nolinkurl{http://ascl.net/#1}}}
\providecommand{\doarXiv}[1]{\href{https://arxiv.org/abs/#1}{\nolinkurl{https://arxiv.org/abs/#1}}}

\bibitem[{{Bannister} {et~al.}(2017){Bannister}, {Shannon}, {Macquart},
  {Flynn}, {Edwards}, {O'Neill}, {Os{\l}owski}, {Bailes}, {Zackay}, {Clarke},
  {D'Addario}, {Dodson}, {Hall}, {Jameson}, {Jones}, {Navarro}, {Trinh},
  {Allison}, {Anderson}, {Bell}, {Chippendale}, {Collier}, {Heald}, {Heywood},
  {Hotan}, {Lee-Waddell}, {Madrid}, {Marvil}, {McConnell}, {Popping},
  {Voronkov}, {Whiting}, {Allen}, {Bock}, {Brodrick}, {Cooray}, {DeBoer},
  {Diamond}, {Ekers}, {Gough}, {Hampson}, {Harvey-Smith}, {Hay}, {Hayman},
  {Jackson}, {Johnston}, {Koribalski}, {McClure-Griffiths}, {Mirtschin}, {Ng},
  {Norris}, {Pearce}, {Phillips}, {Roxby}, {Troup}, \&
  {Westmeier}}]{bannister2017}
{Bannister}, K.~W., {Shannon}, R.~M., {Macquart}, J.~P., {et~al.} 2017, \apjl,
  841, L12, \dodoi{10.3847/2041-8213/aa71ff}

\bibitem[{{Bassa} {et~al.}(2017){Bassa}, {Pleunis}, {Hessels}, {Ferrara},
  {Breton}, {Gusinskaia}, {Kondratiev}, {Sanidas}, {Nieder}, {Clark}, {Li},
  {van Amesfoort}, {Burnett}, {Camilo}, {Michelson}, {Ransom}, {Ray}, \&
  {Wood}}]{bassa2017}
{Bassa}, C.~G., {Pleunis}, Z., {Hessels}, J.~W.~T., {et~al.} 2017, \apjl, 846,
  L20, \dodoi{10.3847/2041-8213/aa8400}

\bibitem[{Bates {et~al.}(2014)Bates, Lorimer, Rane, \& Swiggum}]{Bates2014}
Bates, S.~D., Lorimer, D.~R., Rane, A., \& Swiggum, J. 2014, \mnras, 439, 2893,
  \dodoi{10.1093/mnras/stu157}

\bibitem[{{Bhat} {et~al.}(2004){Bhat}, {Cordes}, {Camilo}, {Nice}, \&
  {Lorimer}}]{bhat2004}
{Bhat}, N.~D.~R., {Cordes}, J.~M., {Camilo}, F., {Nice}, D.~J., \& {Lorimer},
  D.~R. 2004, \apj, 605, 759, \dodoi{10.1086/382680}

\bibitem[{{Bhat} {et~al.}(2016){Bhat}, {Ord}, {Tremblay}, {McSweeney}, \&
  {Tingay}}]{bhat2016}
{Bhat}, N.~D.~R., {Ord}, S.~M., {Tremblay}, S.~E., {McSweeney}, S.~J., \&
  {Tingay}, S.~J. 2016, \apj, 818, 86, \dodoi{10.3847/0004-637X/818/1/86}

\bibitem[{{Bhat} {et~al.}(2018){Bhat}, {Tremblay}, {Kirsten}, {Meyers},
  {Sokolowski}, {van Straten}, {McSweeney}, {Ord}, {Shannon}, {Beardsley},
  {Crosse}, {Emrich}, {Franzen}, {Horsley}, {Johnston-Hollitt}, {Kaplan},
  {Kenney}, {Morales}, {Pallot}, {Steele}, {Tingay}, {Trott}, {Walker},
  {Wayth}, {Williams}, \& {Wu}}]{bhat2018}
{Bhat}, N.~D.~R., {Tremblay}, S.~E., {Kirsten}, F., {et~al.} 2018, \apjs, 238,
  1, \dodoi{10.3847/1538-4365/aad37c}

\bibitem[{{Bhattacharyya} {et~al.}(2016){Bhattacharyya}, {Cooper}, {Malenta},
  {Roy}, {Chengalur}, {Keith}, {Kudale}, {McLaughlin}, {Ransom}, {Ray}, \&
  {Stappers}}]{bhaswati2016}
{Bhattacharyya}, B., {Cooper}, S., {Malenta}, M., {et~al.} 2016, \apj, 817,
  130, \dodoi{10.3847/0004-637X/817/2/130}

\bibitem[{Brentjens \& de~Bruyn(2005)}]{Brentjens2005}
Brentjens, M., \& de~Bruyn, A. 2005, Astronomy {\&} Astrophysics, 441, 1217,
  \dodoi{10.1051/0004-6361:20052990}

\bibitem[{{Cordes} \& {Lazio}(2002)}]{ne2001}
{Cordes}, J.~M., \& {Lazio}, T.~J.~W. 2002, arXiv e-prints, astro.
\newblock \doarXiv{astro-ph/0207156}

\bibitem[{{Cordes} {et~al.}(1985){Cordes}, {Weisberg}, \& {Boriakoff}}]{cwb85}
{Cordes}, J.~M., {Weisberg}, J.~M., \& {Boriakoff}, V. 1985, \apj, 288, 221,
  \dodoi{10.1086/162784}

\bibitem[{{Cordes} {et~al.}(2006){Cordes}, {Freire}, {Lorimer}, {Camilo},
  {Champion}, {Nice}, {Ramachand ran}, {Hessels}, {Vlemmings}, {van Leeuwen},
  {Ransom}, {Bhat}, {Arzoumanian}, {McLaughlin}, {Kaspi}, {Kasian}, {Deneva},
  {Reid}, {Chatterjee}, {Han}, {Backer}, {Stairs}, {Deshpand e}, \&
  {Faucher-Gigu{\`e}re}}]{cordes2006}
{Cordes}, J.~M., {Freire}, P.~C.~C., {Lorimer}, D.~R., {et~al.} 2006, \apj,
  637, 446, \dodoi{10.1086/498335}

\bibitem[{{Dai} {et~al.}(2019){Dai}, {Lower}, {Bailes}, {Camilo}, {Halpern},
  {Johnston}, {Kerr}, {Reynolds}, {Sarkissian}, \& {Scholz}}]{dai2019}
{Dai}, S., {Lower}, M.~E., {Bailes}, M., {et~al.} 2019, \apjl, 874, L14,
  \dodoi{10.3847/2041-8213/ab0e7a}

\bibitem[{{Good} {et~al.}(2020){Good}, {Andersen}, {Chawla}, {Crowter}, {Dong},
  {Fonseca}, {Meyers}, {Ng}, {Pleunis}, {Ransom}, {Stairs}, {Tan}, {Bhardwaj},
  {Boyle}, {Dobbs}, {Gaensler}, {Kaspi}, {Masui}, {Naidu}, {Rafiei-Ravandi},
  {Scholz}, {Smith}, \& {Tendulkar}}]{good2020}
{Good}, D.~C., {Andersen}, B.~C., {Chawla}, P., {et~al.} 2020, arXiv e-prints,
  arXiv:2012.02320.
\newblock \doarXiv{2012.02320}

\bibitem[{{Gupta} {et~al.}(2017){Gupta}, {Ajithkumar}, {Kale}, {Nayak},
  {Sabhapathy}, {Sureshkumar}, {Swami}, {Chengalur}, {Ghosh},
  {Ishwara-Chandra}, {Joshi}, {Kanekar}, {Lal}, \& {Roy}}]{ugmrt}
{Gupta}, Y., {Ajithkumar}, B., {Kale}, H.~S., {et~al.} 2017, Current Science,
  113, 707

\bibitem[{{Hobbs} {et~al.}(2020){Hobbs}, {Manchester}, {Dunning}, {Jameson},
  {Roberts}, {George}, {Green}, {Tuthill}, {Toomey}, {Kaczmarek}, {Mader},
  {Marquarding}, {Ahmed}, {Amy}, {Bailes}, {Beresford}, {Bhat}, {Bock},
  {Bourne}, {Bowen}, {Brothers}, {Cameron}, {Carretti}, {Carter}, {Castillo},
  {Chekkala}, {Cheng}, {Chung}, {Craig}, {Dai}, {Dawson}, {Dempsey}, {Doherty},
  {Dong}, {Edwards}, {Ergesh}, {Gao}, {Han}, {Hayman}, {Indermuehle},
  {Jeganathan}, {Johnston}, {Kanoniuk}, {Kesteven}, {Kramer}, {Leach},
  {Mcintyre}, {Moss}, {Os{\l}owski}, {Phillips}, {Pope}, {Preisig}, {Price},
  {Reeves}, {Reilly}, {Reynolds}, {Robishaw}, {Roush}, {Ruckley}, {Sadler},
  {Sarkissian}, {Severs}, {Shannon}, {Smart}, {Smith}, {Smith}, {Sobey},
  {Staveley-Smith}, {Tzioumis}, {van Straten}, {Wang}, {Wen}, \&
  {Whiting}}]{hobbs2020}
{Hobbs}, G., {Manchester}, R.~N., {Dunning}, A., {et~al.} 2020, \pasa, 37,
  e012, \dodoi{10.1017/pasa.2020.2}

\bibitem[{{Hotan} {et~al.}(2004){Hotan}, {van Straten}, \&
  {Manchester}}]{psrchive}
{Hotan}, A.~W., {van Straten}, W., \& {Manchester}, R.~N. 2004, \pasa, 21, 302,
  \dodoi{10.1071/AS04022}

\bibitem[{{Hurley-Walker} {et~al.}(2017){Hurley-Walker}, {Callingham},
  {Hancock}, {Franzen}, {Hindson}, {Kapi{\'n}ska}, {Morgan}, {Offringa},
  {Wayth}, {Wu}, {Zheng}, {Murphy}, {Bell}, {Dwarakanath}, {For}, {Gaensler},
  {Johnston-Hollitt}, {Lenc}, {Procopio}, {Staveley-Smith}, {Ekers}, {Bowman},
  {Briggs}, {Cappallo}, {Deshpande}, {Greenhill}, {Hazelton}, {Kaplan},
  {Lonsdale}, {McWhirter}, {Mitchell}, {Morales}, {Morgan}, {Oberoi}, {Ord},
  {Prabu}, {Shankar}, {Srivani}, {Subrahmanyan}, {Tingay}, {Webster},
  {Williams}, \& {Williams}}]{gleam2017}
{Hurley-Walker}, N., {Callingham}, J.~R., {Hancock}, P.~J., {et~al.} 2017,
  \mnras, 464, 1146, \dodoi{10.1093/mnras/stw2337}

\bibitem[{{Intema} {et~al.}(2017){Intema}, {Jagannathan}, {Mooley}, \&
  {Frail}}]{tgss2017}
{Intema}, H.~T., {Jagannathan}, P., {Mooley}, K.~P., \& {Frail}, D.~A. 2017,
  \aap, 598, A78, \dodoi{10.1051/0004-6361/201628536}

\bibitem[{{Jankowski} {et~al.}(2018){Jankowski}, {van Straten}, {Keane},
  {Bailes}, {Barr}, {Johnston}, \& {Kerr}}]{fabian2018}
{Jankowski}, F., {van Straten}, W., {Keane}, E.~F., {et~al.} 2018, \mnras, 473,
  4436, \dodoi{10.1093/mnras/stx2476}

\bibitem[{{Kaur} {et~al.}(2019){Kaur}, {Bhat}, {Tremblay}, {Shannon},
  {McSweeney}, {Ord}, {Beardsley}, {Crosse}, {Emrich}, {Franzen}, {Horsley},
  {Johnston-Hollitt}, {Kaplan}, {Kenney}, {Morales}, {Pallot}, {Steele},
  {Tingay}, {Trott}, {Walker}, {Wayth}, {Williams}, \& {Wu}}]{kaur2019}
{Kaur}, D., {Bhat}, N.~D.~R., {Tremblay}, S.~E., {et~al.} 2019, \apj, 882, 133,
  \dodoi{10.3847/1538-4357/ab338f}

\bibitem[{{Keane} {et~al.}(2015){Keane}, {Bhattacharyya}, {Kramer}, {Stappers},
  {Keane}, {Bhattacharyya}, {Kramer}, {Stappers}, {Bates}, {Burgay},
  {Chatterjee}, {Champion}, {Eatough}, {Hessels}, {Janssen}, {Lee}, {van
  Leeuwen}, {Margueron}, {Oertel}, {Possenti}, {Ransom}, {Theureau}, \&
  {Torne}}]{keane2015}
{Keane}, E., {Bhattacharyya}, B., {Kramer}, M., {et~al.} 2015, in Advancing
  Astrophysics with the Square Kilometre Array (AASKA14), 40.
\newblock \doarXiv{1501.00056}

\bibitem[{{Keith} {et~al.}(2010){Keith}, {Jameson}, {van Straten}, {Bailes},
  {Johnston}, {Kramer}, {Possenti}, {Bates}, {Bhat}, {Burgay}, {Burke-Spolaor},
  {D'Amico}, {Levin}, {McMahon}, {Milia}, \& {Stappers}}]{keith2010}
{Keith}, M.~J., {Jameson}, A., {van Straten}, W., {et~al.} 2010, \mnras, 409,
  619, \dodoi{10.1111/j.1365-2966.2010.17325.x}

\bibitem[{{Levin} {et~al.}(2013){Levin}, {Bailes}, {Barsdell}, {Bates}, {Bhat},
  {Burgay}, {Burke-Spolaor}, {Champion}, {Coster}, {D'Amico}, {Jameson},
  {Johnston}, {Keith}, {Kramer}, {Milia}, {Ng}, {Possenti}, {Stappers},
  {Thornton}, \& {van Straten}}]{levin2013}
{Levin}, L., {Bailes}, M., {Barsdell}, B.~R., {et~al.} 2013, \mnras, 434, 1387,
  \dodoi{10.1093/mnras/stt1103}

\bibitem[{{Lorimer} {et~al.}(2006){Lorimer}, {Faulkner}, {Lyne}, {Manchester},
  {Kramer}, {McLaughlin}, {Hobbs}, {Possenti}, {Stairs}, {Camilo}, {Burgay},
  {D'Amico}, {Corongiu}, \& {Crawford}}]{lorimer2006}
{Lorimer}, D.~R., {Faulkner}, A.~J., {Lyne}, A.~G., {et~al.} 2006, \mnras, 372,
  777, \dodoi{10.1111/j.1365-2966.2006.10887.x}

\bibitem[{Manchester {et~al.}(1973)Manchester, Taylor, \&
  Huguenin}]{Manchester1973}
Manchester, R., Taylor, J., \& Huguenin, G. 1973, \apjl, 179, L7,
  \dodoi{10.1086/181105}

\bibitem[{{Manchester} {et~al.}(1996){Manchester}, {Lyne}, {D'Amico}, {Bailes},
  {Johnston}, {Lorimer}, {Harrison}, {Nicastro}, \& {Bell}}]{70cm}
{Manchester}, R.~N., {Lyne}, A.~G., {D'Amico}, N., {et~al.} 1996, \mnras, 279,
  1235, \dodoi{10.1093/mnras/279.4.1235}

\bibitem[{{Manchester} {et~al.}(2001){Manchester}, {Lyne}, {Camilo}, {Bell},
  {Kaspi}, {D'Amico}, {McKay}, {Crawford}, {Stairs}, {Possenti}, {Kramer}, \&
  {Sheppard}}]{manchester2001}
{Manchester}, R.~N., {Lyne}, A.~G., {Camilo}, F., {et~al.} 2001, \mnras, 328,
  17, \dodoi{10.1046/j.1365-8711.2001.04751.x}

\bibitem[{{Maron} {et~al.}(2000){Maron}, {Kijak}, {Kramer}, \&
  {Wielebinski}}]{maron2000}
{Maron}, O., {Kijak}, J., {Kramer}, M., \& {Wielebinski}, R. 2000, \aaps, 147,
  195, \dodoi{10.1051/aas:2000298}

\bibitem[{{McMullin} {et~al.}(2007){McMullin}, {Waters}, {Schiebel}, {Young},
  \& {Golap}}]{casa}
{McMullin}, J.~P., {Waters}, B., {Schiebel}, D., {Young}, W., \& {Golap}, K.
  2007, in Astronomical Society of the Pacific Conference Series, Vol. 376,
  Astronomical Data Analysis Software and Systems XVI, ed. R.~A. {Shaw},
  F.~{Hill}, \& D.~J. {Bell}, 127

\bibitem[{{McSweeney} {et~al.}(2017){McSweeney}, {Bhat}, {Tremblay}, {Deshpand
  e}, \& {Ord}}]{mcsweeney2017}
{McSweeney}, S.~J., {Bhat}, N.~D.~R., {Tremblay}, S.~E., {Deshpand e}, A.~A.,
  \& {Ord}, S.~M. 2017, \apj, 836, 224, \dodoi{10.3847/1538-4357/aa5c35}

\bibitem[{{McSweeney} {et~al.}(2019){McSweeney}, {Bhat}, {Wright}, {Tremblay},
  \& {Kudale}}]{mcsweeney2019}
{McSweeney}, S.~J., {Bhat}, N.~D.~R., {Wright}, G., {Tremblay}, S.~E., \&
  {Kudale}, S. 2019, \apj, 883, 28, \dodoi{10.3847/1538-4357/ab3a97}

\bibitem[{{McSweeney} {et~al.}(2020){McSweeney}, {Ord}, {Kaur}, {Bhat},
  {Meyers}, {Tremblay}, {Jones}, {Crosse}, \& {Smith}}]{McSweeney2020}
{McSweeney}, S.~J., {Ord}, S.~M., {Kaur}, D., {et~al.} 2020, \pasa, 37, e034,
  \dodoi{10.1017/pasa.2020.24}

\bibitem[{{Meyers} {et~al.}(2018){Meyers}, {Tremblay}, {Bhat}, {Flynn},
  {Gupta}, {Shannon}, {Murray}, {Sobey}, {Ord}, {Os{\l}owski}, {Crosse},
  {Williams}, {Jankowski}, {Farah}, {Venkatraman Krishnan}, {Bateman},
  {Bailes}, {Beardsley}, {Emrich}, {Franzen}, {Gaensler}, {Horsley},
  {Johnston-Hollitt}, {Kaplan}, {Kenney}, {Morales}, {Pallot}, {Steele},
  {Tingay}, {Trott}, {Walker}, {Wayth}, \& {Wu}}]{meyers2018}
{Meyers}, B.~W., {Tremblay}, S.~E., {Bhat}, N.~D.~R., {et~al.} 2018, \apj, 869,
  134, \dodoi{10.3847/1538-4357/aaee7b}

\bibitem[{{Mohan} \& {Rafferty}(2015)}]{pybdsm}
{Mohan}, N., \& {Rafferty}, D. 2015, {PyBDSF: Python Blob Detection and Source
  Finder}.
\newblock \doeprint{1502.007}

\bibitem[{{Ord} {et~al.}(2019){Ord}, {Tremblay}, {McSweeney}, {Bhat}, {Sobey},
  {Mitchell}, {Hancock}, \& {Kirsten}}]{Ord2019}
{Ord}, S.~M., {Tremblay}, S.~E., {McSweeney}, S.~J., {et~al.} 2019, \pasa, 36,
  e030, \dodoi{10.1017/pasa.2019.17}

\bibitem[{{Prasad} \& {Chengalur}(2012)}]{flagcal}
{Prasad}, J., \& {Chengalur}, J. 2012, Experimental Astronomy, 33, 157,
  \dodoi{10.1007/s10686-011-9279-5}

\bibitem[{Radhakrishnan \& Cooke(1969)}]{Radhakrishnan1969}
Radhakrishnan, V., \& Cooke, D.~J. 1969, Astrophysical Letters, 3, 225

\bibitem[{{Ransom}(2001)}]{presto}
{Ransom}, S.~M. 2001, PhD thesis, Harvard University

\bibitem[{{Reddy} {et~al.}(2017){Reddy}, {Kudale}, {Gokhale}, {Halagalli},
  {Raskar}, {de}, {Gnanaraj}, {Ajith Kumar}, \& {Gupta}}]{gwb}
{Reddy}, S.~H., {Kudale}, S., {Gokhale}, U., {et~al.} 2017, Journal of
  Astronomical Instrumentation, 6, 1641011, \dodoi{10.1142/S2251171716410117}

\bibitem[{{Sanidas} {et~al.}(2019){Sanidas}, {Cooper}, {Bassa}, {Hessels},
  {Kondratiev}, {Michilli}, {Stappers}, {Tan}, {van Leeuwen}, {Cerrigone},
  {Fallows}, {Iacobelli}, {Orr{\'u}}, {Pizzo}, {Shulevski}, {Toribio}, {ter
  Veen}, {Zucca}, {Bondonneau}, {Grie{\ss}meier}, {Karastergiou}, {Kramer}, \&
  {Sobey}}]{sanidas2019}
{Sanidas}, S., {Cooper}, S., {Bassa}, C.~G., {et~al.} 2019, \aap, 626, A104,
  \dodoi{10.1051/0004-6361/201935609}

\bibitem[{{Shannon} {et~al.}(2018){Shannon}, {Macquart}, {Bannister}, {Ekers},
  {James}, {Os{\l}owski}, {Qiu}, {Sammons}, {Hotan}, {Voronkov}, {Beresford},
  {Brothers}, {Brown}, {Bunton}, {Chippendale}, {Haskins}, {Leach},
  {Marquarding}, {McConnell}, {Pilawa}, {Sadler}, {Troup}, {Tuthill},
  {Whiting}, {Allison}, {Anderson}, {Bell}, {Collier}, {G{\"u}rkan}, {Heald},
  \& {Riseley}}]{shannon2018}
{Shannon}, R.~M., {Macquart}, J.~P., {Bannister}, K.~W., {et~al.} 2018, \nat,
  562, 386, \dodoi{10.1038/s41586-018-0588-y}

\bibitem[{{Stovall} {et~al.}(2014){Stovall}, {Lynch}, {Ransom}, {Archibald},
  {Banaszak}, {Biwer}, {Boyles}, {Dartez}, {Day}, {Ford}, {Flanigan}, {Garcia},
  {Hessels}, {Hinojosa}, {Jenet}, {Kaplan}, {Karako-Argaman}, {Kaspi},
  {Kondratiev}, {Leake}, {Lorimer}, {Lunsford}, {Martinez}, {Mata},
  {McLaughlin}, {Roberts}, {Rohr}, {Siemens}, {Stairs}, {van Leeuwen},
  {Walker}, \& {Wells}}]{stovall2014}
{Stovall}, K., {Lynch}, R.~S., {Ransom}, S.~M., {et~al.} 2014, \apj, 791, 67,
  \dodoi{10.1088/0004-637X/791/1/67}

\bibitem[{{Tan} {et~al.}(2018{\natexlab{a}}){Tan}, {Lyon}, {Stappers},
  {Cooper}, {Hessels}, {Kondratiev}, {Michilli}, \& {Sanidas}}]{tan2018a}
{Tan}, C.~M., {Lyon}, R.~J., {Stappers}, B.~W., {et~al.} 2018{\natexlab{a}},
  \mnras, 474, 4571, \dodoi{10.1093/mnras/stx3047}

\bibitem[{{Tan} {et~al.}(2018{\natexlab{b}}){Tan}, {Bassa}, {Cooper},
  {Dijkema}, {Esposito}, {Hessels}, {Kondratiev}, {Kramer}, {Michilli},
  {Sanidas}, {Shimwell}, {Stappers}, {van Leeuwen}, {Cognard},
  {Grie{\ss}meier}, {Karastergiou}, {Keane}, {Sobey}, \&
  {Weltevrede}}]{tan2018}
{Tan}, C.~M., {Bassa}, C.~G., {Cooper}, S., {et~al.} 2018{\natexlab{b}}, \apj,
  866, 54, \dodoi{10.3847/1538-4357/aade88}

\bibitem[{Tan {et~al.}(2020)Tan, Bassa, Cooper, Hessels, Kondratiev, Michilli,
  Sanidas, Stappers, van Leeuwen, Donner, Grießmeier, Kramer, Tiburzi,
  Weltevrede, Ciardi, Hoeft, Mann, Miskolczi, Schwarz, Vocks, \&
  Wucknitz}]{Tan2020}
Tan, C.~M., Bassa, C.~G., Cooper, S., {et~al.} 2020, \mnras, 492, 5878,
  \dodoi{10.1093/mnras/staa113}

\bibitem[{{Tingay} {et~al.}(2013){Tingay}, {Goeke}, {Bowman}, {Emrich}, {Ord},
  {Mitchell}, {Morales}, {Booler}, {Crosse}, {Wayth}, {Lonsdale}, {Tremblay},
  {Pallot}, {Colegate}, {Wicenec}, {Kudryavtseva}, {Arcus}, {Barnes},
  {Bernardi}, {Briggs}, {Burns}, {Bunton}, {Cappallo}, {Corey}, {Deshpande},
  {Desouza}, {Gaensler}, {Greenhill}, {Hall}, {Hazelton}, {Herne}, {Hewitt},
  {Johnston-Hollitt}, {Kaplan}, {Kasper}, {Kincaid}, {Koenig}, {Kratzenberg},
  {Lynch}, {Mckinley}, {Mcwhirter}, {Morgan}, {Oberoi}, {Pathikulangara},
  {Prabu}, {Remillard}, {Rogers}, {Roshi}, {Salah}, {Sault}, {Udaya-Shankar},
  {Schlagenhaufer}, {Srivani}, {Stevens}, {Subrahmanyan}, {Waterson},
  {Webster}, {Whitney}, {Williams}, {Williams}, \& {Wyithe}}]{tingay2013}
{Tingay}, S.~J., {Goeke}, R., {Bowman}, J.~D., {et~al.} 2013, \pasa, 30, e007,
  \dodoi{10.1017/pasa.2012.007}

\bibitem[{{Tremblay} {et~al.}(2015){Tremblay}, {Ord}, {Bhat}, {Tingay},
  {Crosse}, {Pallot}, {Oronsaye}, {Bernardi}, {Bowman}, {Briggs}, {Cappallo},
  {Corey}, {Deshpand e}, {Emrich}, {Goeke}, {Greenhill}, {Hazelton},
  {Johnston-Hollitt}, {Kaplan}, {Kasper}, {Kratzenberg}, {Lonsdale}, {Lynch},
  {McWhirter}, {Mitchell}, {Morales}, {Morgan}, {Oberoi}, {Prabu}, {Rogers},
  {Roshi}, {Udaya Shankar}, {Srivani}, {Subrahmanyan}, {Waterson}, {Wayth},
  {Webster}, {Whitney}, {Williams}, \& {Williams}}]{Tremblay2015}
{Tremblay}, S.~E., {Ord}, S.~M., {Bhat}, N.~D.~R., {et~al.} 2015, \pasa, 32,
  e005, \dodoi{10.1017/pasa.2015.6}

\bibitem[{{van Straten} \& {Bailes}(2011)}]{dspsr}
{van Straten}, W., \& {Bailes}, M. 2011, \pasa, 28, 1, \dodoi{10.1071/AS10021}

\bibitem[{{Wayth} {et~al.}(2018){Wayth}, {Tingay}, {Trott}, {Emrich},
  {Johnston-Hollitt}, {McKinley}, {Gaensler}, {Beardsley}, {Booler}, {Crosse},
  {Franzen}, {Horsley}, {Kaplan}, {Kenney}, {Morales}, {Pallot}, {Sleap},
  {Steele}, {Walker}, {Williams}, {Wu}, {Cairns}, {Filipovic}, {Johnston},
  {Murphy}, {Quinn}, {Staveley-Smith}, {Webster}, \& {Wyithe}}]{wayth2018}
{Wayth}, R.~B., {Tingay}, S.~J., {Trott}, C.~M., {et~al.} 2018, \pasa, 35, 33,
  \dodoi{10.1017/pasa.2018.37}

\bibitem[{Xue(2019)}]{Xue2019a}
Xue, M. 2019, PhD thesis

\bibitem[{Xue {et~al.}(2019)Xue, Ord, Tremblay, Bhat, Sobey, Meyers, McSweeney,
  \& Swainston}]{Xue2019}
Xue, M., Ord, S.~M., Tremblay, S.~E., {et~al.} 2019, Publications of the
  Astronomical Society of Australia, 36, e025, \dodoi{10.1017/pasa.2019.19}

\bibitem[{{Xue} {et~al.}(2017){Xue}, {Bhat}, {Tremblay}, {Ord}, {Sobey},
  {Swainston}, {Kaplan}, {Johnston}, {Meyers}, \& {McSweeney}}]{xue2017}
{Xue}, M., {Bhat}, N.~D.~R., {Tremblay}, S.~E., {et~al.} 2017, \pasa, 34, e070,
  \dodoi{10.1017/pasa.2017.66}

\bibitem[{{Yao} {et~al.}(2017){Yao}, {Manchester}, \& {Wang}}]{ymw16}
{Yao}, J.~M., {Manchester}, R.~N., \& {Wang}, N. 2017, \apj, 835, 29,
  \dodoi{10.3847/1538-4357/835/1/29}

\end{thebibliography}
\bibliographystyle{aasjournal}

\end{document}